\documentclass[11pt]{article}

\usepackage[english]{babel}
\usepackage{graphicx}
\usepackage{latexsym}
\usepackage{amssymb}
\usepackage{amsmath}
\usepackage{float}
\usepackage{mismath}
\usepackage{xcolor}
\usepackage{hyperref}

\begin{document}

\title{Kantowski-Sachs cosmological model with axion-like scalar field and dark energy 
}
\author{%
    Maxim Krasnov $^{1}$\thanks{morrowindman1@mail.ru}, 
    Oem Trivedi$^{2}$\thanks{oem.t@ahduni.edu.in}, 
    Maxim Khlopov $^{1,3,4}$\thanks{khlopov@apc.in2p3.fr}
}
\date{%
    \small
    $^{1}$Center for Cosmoparticle Physics Cosmion, National Research Nuclear University “MEPHI”, 115409 Moscow, Russia \\
    $^{2}$International Centre for Space Sciences and Cosmology, Ahmedabad University, Ahmedabad 380009, India\\
    $^{3}$Research Institute of Physics, Southern Federal University, 344090 Rostov-on-Don, Russia\\
    $^{4}$Virtual Institute of Astroparticle Physics, 75018 Paris, France\\    
    \today 
}
\maketitle

\begin{abstract}
   \textcolor{black}{We consider Axion-like particle (ALP) model to construct numerical spatially homogeneous anisotropic Kantowski-Sachs cosmological model. We present various analytical and numerical results in this regard, discussing the evolution of various important cosmological parameters in this regard with ALP scalar field. We also present some general results for generic scalar field in a Kantowski-Sachs background.}
\end{abstract}


\section{Introduction}
\textcolor{black}{Axion-like models are popular dark matter candidates and offer intriguing possibilities for unifying particle physics and cosmology.} The ongoing quest to identify the constituents of dark matter has spurred investigations into a plethora of hypothetical particles beyond the Standard Model (SM). Originally envisioned as an extension of the Peccei-Quinn (PQ) mechanism \cite{peccei1977cp} to address the strong CP problem in Quantum Chromodynamics (QCD), these pseudo-Nambu-Goldstone bosons have been postulated in various frameworks encompassing quantum gravity \cite{arvanitaki2010string,svrcek2006axions,conlon2006qcd}. ALPs have garnered significant interest as potential dark matter constituents as despite their inherently feeble masses ($m_a \lesssim 1$ keV), non-thermal production mechanisms in the early universe could have yielded a population of ALPs that persists today, potentially constituting the bulk of cold dark matter (CDM). The appeal of ALPs as dark matter candidates lies in their ability to circumvent the limitations of the standard freeze-out mechanism, which plagues weakly interacting massive particles WIMPs in their dark matter candidature. The low mass of ALPs allows them to remain in thermal equilibrium with the bath of particles in the early universe for an extended period and this period evades the issue of WIMPs becoming too sluggish to interact efficiently after freeze-out, leading to an underabundance of relic WIMPs compared to the observed dark matter density \cite{Buschmann:2019icd,Iwazaki:1997bk,gorghetto2023post,Bernal:2021yyb}.
\\
\\
Exploring the theoretical status and experimental constraints on ALP properties is crucial for elucidating their potential role in cosmology. In this regard, the recent advancements in the direct observation of gravitational waves (GWs) are driving substantial progress \cite{LIGOScientific:2016aoc,LIGOScientific:2016sjg,Maggiore:2007ulw,Maggiore:1999vm}. Direct observations of GWs have long been expected to yield crucial insights into physics at extremely high energies due to their weak interaction with matter, allowing them to preserve nearly all characteristics of astrophysical or cosmological events \cite{Rubin:2000dq,Khlopov:2004sc,Garriga:2015fdk,Deng:2016vzb,Liu:2022bvr}. Over the past decades, experimental sensitivities for the direct detection of GWs have significantly improved and numerous new GW observatories are planned worldwide and within this framework, it is imperative to explore various potential sources of GWs and determine the extent to which new physics can be inferred from future observations \cite{Grishchuk:1974ny,Starobinsky:1979ty,Witten:1984rs,Kamionkowski:1993fg,Vilenkin:1981bx,Accetta:1988bg,Khlebnikov:1997di,Easther:2006gt,Garcia-Bellido:2007fiu}. Among the possible cosmological sources of GWs are topological defects in the early universe, such as cosmic strings and domain walls \cite{Kibble:1976sj,Vilenkin:1981zs}. Domain walls are sheet-like topological defects that could form in the early universe when a discrete symmetry is spontaneously broken. Given that discrete symmetries are pervasive in high-energy physics beyond the Standard Model (SM), many new physics models predict the formation of domain walls in the early universe. By examining their cosmological evolution, we can derive several constraints on these models, even if their energy scales exceed those probed by laboratory experiments. Many models of ALP fields, for example, have been considered with regards to creation of domain walls in the early universe \cite{Dunsky:2024zdo,Blasi:2023sej,Hiramatsu:2012sc,Nayak:2000pf,dine2021comments,Blasi:2022ayo}. 
\\
\\
Typically the formation of domain walls is considered problematic in cosmology, as their energy density can quickly dominate the total energy density of the universe which one might take as a contradiction to current observational data. However, there is the possibility that domain walls are unstable and collapse before they close the universe and their instability might be ensured if the discrete symmetry is only approximate and explicitly broken by a small parameter in the theory. In such a scenario a significant amount of GWs can be produced during the collisions and annihilation of domain walls, potentially persisting as a stochastic GW background (SGWB) in the present universe. Observations of these relic GWs would enable us to trace events in the very early universe and provide a novel method for investigating physics at extremely high energies. It was recently shown that ALP fields can lead to the formation of closed domain walls given an inflationary universe scenario, which could be reflected in the origin of the nHz SGWB discovered by PTA facilities, and of the early galaxy formation indicated by the James Webb Space Telescope (JWST) data \cite{Guo:2023hyp}. \textcolor{black}{ In the next section, we discuss in brief about some general properties of the evolution of ALP fields in the early universe, while in section III  we derive some important analytical results and numerically analyze Kantowski-Sachs cosmological model with ALP scalar field and cosmic fluid, after that we present general Kantowski-Sachs model for generic scalar field. We conclude our work in section IV, with some discussion on the implications of the results we have obtained.}
\section{Evolution of the ALP field in the inflationary Universe}
ALPs can display some very interesting in the early universe. At the inflationary stage, characterized by the Hubble parameter $H_i$, the amplitude of the complex field $\Psi$ acquires a vacuum expectation value $f$, while its phase $\theta_0$ is set to a value less than $\pi$ at the e-folding, corresponding to the scale of the modern cosmological horizon. During subsequent stages of inflation, the phase fluctuates on small scales and can cross $\pi$, generating contours of future closed walls that separate the vacua of the ALP field, specifically $\theta_v = 0$ and $\theta_v = 2\pi$. After reheating, vacuum domain walls with surface density $\sigma = f \Lambda^2$ are formed when the Hubble parameter $H_w$ becomes equal to the mass of the ALP particle, $m = \Lambda^2 / f$.
\\
\\
For detailed references on the behaviour of ALPs in the early universe and their relations with domain walls and gravitational waves, we would refer the reader to \cite{Marsh:2015xka,Ballesteros:2016xej,Steinhardt:1983ia,Flacke:2016szy,Ballesteros:2016euj,Espinosa:2015eda,Choi:2020rgn,Niemeyer:2019aqm,Daido:2017wwb,Thorne:2017jft,Zhou:2020kkf,Odintsov:2020nwm,Barnaby:2009dd,Cicoli:2023opf,Ferreira:2014zia,Gorghetto:2021fsn,Daido:2017tbr,Co:2021lkc,Domcke:2019mnd,Cheng:2015oqa,Giovannini:1999by,Oikonomou:2023qfz,Choi:2016luu,Visinelli:2017imh,Tashiro:2013yea,Xiao:2021nkb,Hardy:2016mns}. The energy density within the wall is $\Lambda^4$, and the width of the wall is approximately, $1/m = f / \Lambda^2$, and if a wall is sufficiently large, it starts to dominate within the horizon at $$t_d = 1/H_d = m_P^2 / (\Lambda^2 f)$$ before it entirely enters the horizon. Assuming a radiation-dominated stage post-reheating, the temperature of the plasma during the period of wall formation is given by $$T_w = \sqrt{H_w m_P} = \Lambda \sqrt{m_P / f}$$ while wall dominance within the horizon occurs at a plasma temperature of $$T_d = \Lambda \sqrt{f / m_P}$$ During the period of wall dominance, the mass of the relativistic plasma within the horizon is given by $$M_r = m_P^4 / (\Lambda^2 f)$$ The energy density of ALP field oscillations is redshifted after the period of wall formation by the cube of the ratio of scale factors $a_w / a_d = T_d / T_w = f / m_P$, resulting in an ALP mass within the horizon of $$M_a = m_P^3 / \Lambda^2 = (f / m_P) M_r \ll M_r$$ Consequently, the evolution of the universe region where the wall dominates is governed by a wall-dominated, non-homogeneous, and anisotropic spacetime, in which the plasma and matter (ALP field oscillations) play a negligible dynamical role.
\\
\\
It thus makes it possible for one to consider first the evolution of the space-time in which thin surface of false vacuum - domain wall dominates in the dynamics of expansion, and then to sew the obtained solution with the surrounding FRW part of the Universe and consider evolution of ALP field surrounding the wall dominated region. It should also be noted that in the period of wall formation $H_w = m$, so that the width of wall is comparable with the Hubble radius, but these thick walls don't play dominant dynamical role. When walls start to dominate, $$H_d =  (\Lambda^2 f)/m_P^2= (f/m_P)^2 m \ll m$$ which means that the wall width is much less than the Hubble radius and the thin wall approximation is appropriate for our task. 
\section{ALP cosmological model for a distant observer and generic scalar analysis }
\begin{figure}[ht]
	\begin{center}
\includegraphics[width=0.9\textwidth]{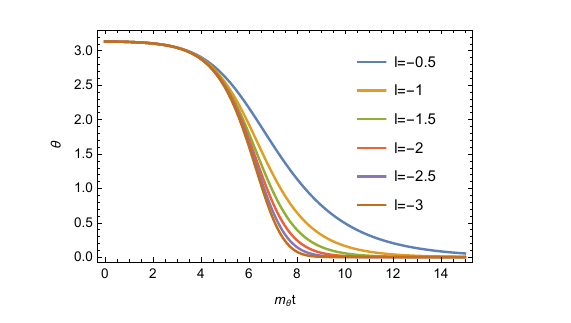}
	\end{center}
\caption{The plot of the solution of \eqref{thetaEq} for different values of of $l$, where the interesting observation is the impact of $l$ changing slowly for $l<-2$.}
	\label{imp_of_l}
\end{figure}

\begin{figure}[H]
	\begin{center}
\includegraphics[width=0.9\textwidth]{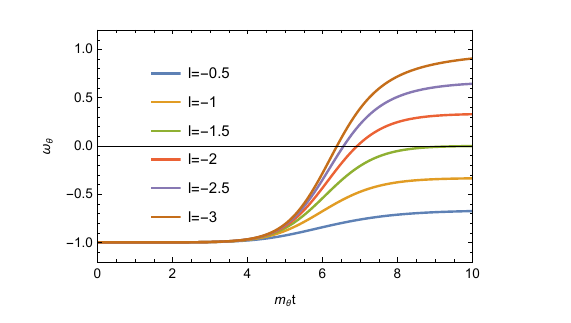}
	\end{center}
\caption{\textcolor{black}{The plot of the equation of state parameter for the scalar field. We see that scalar field could behave differently depending on the value of $l$ and in particular, if $l=-1.5$ then scalar field behaves like non-relativistic matter.}}	\label{imp_of_l_omega_theta}
\end{figure}

\begin{figure}[H]
	\begin{center}
\includegraphics[width=0.9\textwidth]{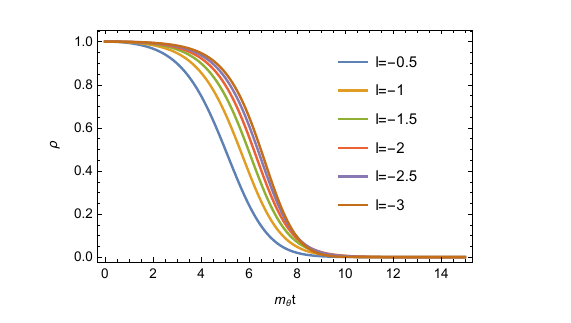}
	\end{center}
\caption{\textcolor{black}{The graph of the fluid's energy density. For any case, when scalar expansion approaches constant value, eventually so do energy density. In case $l=3$ there would be a moment, when energy density would start to increase, since scalar expansion grows infinitely for $l=3$.}}
	\label{imp_of_l_density}
\end{figure}

\textcolor{black}{We are interested in this work towards general space-times, which means we are allowing for inhomogeneity and anisotropic features in our cosmological background as well.} We start by considering a space-time with line element given by 
\begin{equation}\label{InitialLineEl}
    ds^2=A(t,r)^2dt^2 - X(t,r)^2dr^2-Y(t,r)^2d\Omega^2,
\end{equation}
where $d\Omega^2 = d\theta^2+\sin^2{\theta}d\phi^2$. \textcolor{black}{One may notice abuse of symbols in our paper, we denote axion-like scalar field with both $\theta$ and $\phi$ symbols as well as angular coordinates. But there is no explicit utilization of angular coordinates in our analysis, thus, there would be no misunderstanding.  }
Following \cite{Deng:2016vzb}, we set $A(t,r) = 1$, corresponding to gauge symmetry.
ALP Lagrangian is set to be
\begin{equation}\label{lagr}
    \mathcal{L} = \cfrac{1}{2}\,g^{\mu\nu}\partial_\mu\phi \partial_\nu\phi - \Lambda^4 \left[1-\cos{\left(\frac{\phi}{f} \right)}\right]. 
\end{equation}
We also take into account anisotropy of the space time via energy-momentum tensor of cosmological fluid:
\textcolor{black}{
\begin{equation}\label{EMT}
     T^\mu_\nu=\text{diag}[1, \omega, -(\omega+\delta), -(\omega+\gamma)]\rho,
\end{equation}
where $\omega=p/\rho$.
}
\textcolor{black}{As a first step in our analysis we consider limit $r\rightarrow \infty$, i.e. what a distant observer could detect. In that case line element \eqref{InitialLineEl} could be rewritten as
\begin{equation}\label{KSmetric}
    ds^2=dt^2 - X(t)^2dr^2-Y(t)^2d\Omega^2,
\end{equation}
which is Kantowski-Sachs space time \cite{kantowski1966some}. Here we assume that $Y$ is finite as $r\rightarrow \infty$. We also set common border condition for the scalar field $\phi_r(t, \infty) = 0.$ 
\\
\\
Now let us write down system of Einstein's field equations using line element \eqref{KSmetric}, Lagrangian \eqref{lagr} and cosmological fluid' energy-momentum tensor \eqref{EMT}:}

\begin{align}
    \frac{2 \Dot{X} \Dot{Y}}{X Y}+\frac{\Dot{Y}^2}{Y^2}+\frac{1}{Y^2} &= \cfrac{1}{2}\,\Dot{\phi}^2+\Lambda^4\left[ 1-\cos{\left(\cfrac{\phi}{f}\right)}\right] + \rho, \label{eq1}\\
    \frac{2 \Ddot{Y}}{Y}+\frac{\Dot{Y}^2}{Y^2}+\frac{1}{Y^2} &= -\cfrac{1}{2}\,\Dot{\phi}^2+\Lambda^4\left[ 1-\cos{\left(\cfrac{\phi}{f}\right)}\right] -\omega\rho, \label{eq2}\\
    \frac{\Ddot{X}}{X}+\frac{\Dot{X} \Dot{Y}}{X Y}+\frac{\Ddot{Y}}{Y} &= -\cfrac{1}{2}\,\Dot{\phi}^2+\Lambda^4\left[ 1-\cos{\left(\cfrac{\phi}{f}\right)}\right] -(\omega+\delta)\rho, \label{eq3}\\
    \frac{\Ddot{X}}{X}+\frac{\Dot{X} \Dot{Y}}{X Y}+\frac{\Ddot{Y}}{Y} &= -\cfrac{1}{2}\,\Dot{\phi}^2+\Lambda^4\left[ 1-\cos{\left(\cfrac{\phi}{f}\right)}\right] -(\omega+\gamma)\rho. \label{eq4}
\end{align}
Equation of motion of the scalar field is as follows:
\begin{equation}\label{kgeq}
    \Ddot{\phi}+\left( \cfrac{\Dot{X}}{X}+2\,\cfrac{\Dot{Y}}{Y}\right)\Dot{\phi}+\cfrac{\Lambda^4}{f}\,\sin{\left( \cfrac{\phi}{f} \right)}=0.
\end{equation}
\textcolor{black}{Equations \eqref{eq3} and \eqref{eq4} immediately yield $\gamma = \delta$ and now we are given with four independent equations and six variables. We need to make two assumptions to make system solvable.}
\\
\\
To make progress now, we assume the following relation between $X$ and $Y$: 
\begin{equation}\label{AnisotSol}
    Y=X^n,
\end{equation}
\textcolor{black}{which has been commonly utlized in some previous works \cite{Collins, LRS, BianchiVih, Tiwari}.} Furthermore, it is usually assumed that scalar field is proportional to average scale factor taken in some power \cite{johri1994cosmological, johri1989bd, singh2012frw} 
\begin{equation}\label{fieldandscale}
    \phi \propto a(t)^l,
\end{equation}
Average scale factor $a(t)$ in Kantowski-Sachs space-time is defined as
\begin{equation}\label{AvScFac}
    a(t)^3=XY^2.
\end{equation}
\textcolor{black}{One can then check that} relations \eqref{fieldandscale} and \eqref{AvScFac}, \eqref{AnisotSol} leads to the following relation between the scalar field and metric's potential
\begin{equation}\label{assumption}
    \cfrac{\Dot{\phi}}{\phi}=\cfrac{l}{3}\,(2n+1)\,\cfrac{\Dot{X}}{X}.
\end{equation}
We now utilize \eqref{assumption} in \eqref{kgeq}, which leads us to arrive at
\begin{equation}\label{maineq}
    \Ddot{\phi}+\cfrac{3\Dot{\phi}^2}{l\phi}+\cfrac{\Lambda^4}{f}\,\sin{\left( \cfrac{\phi}{f} \right)}=0.
\end{equation}
\textcolor{black}{Let us now perform the following variable substitution $\phi = f\theta$ and switch from time derivative to the derivative with respect to $m_\theta t=t\Lambda^2/f$ (represented by prime). We obtain} 

\begin{equation}\label{thetaEq}
    \theta'' + \cfrac{3}{l}\,\cfrac{\theta'^2}{\theta}+\sin{\theta}=0.
\end{equation}
At this point, we could refer to $\theta$ as a phase and we could then plot the numerical solution of \eqref{thetaEq} for different values of $l$. We set initial conditions for $\theta$ as follows
\begin{equation}
    \theta_{in}=\pi-0.01,\,\,\theta'_{in}=0.
\end{equation}
\textcolor{black}{This allows us to see the evolution of the scalar field, which we have shown in Figure \ref{imp_of_l}.}
\textcolor{black}{The equation of state parameter, given by the usual definition $$\omega_\theta = \frac{p_\theta}{\rho_\theta}$$ is plotted in Fig.\eqref{imp_of_l_omega_theta}} \footnote{\textcolor{black}{Note that the parameter has also been scaled appropriately with regards to the differential equation with $\theta$ variable.}}
\textcolor{black}{From the Einstein equations \eqref{eq1}-\eqref{eq4}, we find that equation of state parameter for cosmological fluid $\omega \approx \text{const} \sim-1$ and skewness parameter $\delta=\gamma \approx 1$ for any value of $l$.} \textcolor{black}{We feel that it is important to emphasize what happens with this behaviour of the skewness parameters. Note from the Einstein equations \eqref{eq4} and the definition of the energy-momentum tensor \eqref{EMT} that the skewness parameters taking this value leads to the angular pressures of the going to zero eventually. The fact that the pressures in the angular components vanish could suggest here that the ALP's energy density is concentrated primarily along the radial direction while allowing free expansion in angular directions. This can be interpreted as a decoupling of the radial and angular pressure behaviors in the energy-momentum tensor. It could also mean that the ALP field likely maintains its coupling in a way that provides pressure support only along the radial component, which is also a general result in this case as this does not necessarily depend on specific choices in the coupling function in our case. Also the angular pressures' vanishing implies that the angular directions are in a state similar to a "dust-like" fluid behavior in those dimensions, where they are not dynamically significant for an observer which is far away as we are interested in the $r \to \infty$ limit anyways.}
\\
\\
\textcolor{black}{ 
Calculation of equation of state' parameter and skewness parameter are dependent on the initial value of $Y_{in}$, which we set to be $Y_{in}=1$. Another choice of initial condition could affect $\omega, \delta, \rho$, but the changes would not be large enough to affect our main conclusions here. In figure\ref{imp_of_l_density} we have also plotted the evolution of the energy density.}  
One can also plot and calculate for the scalar expansion parameter,jerk parameter, deceleration parameter and the average anisotropy, as 
\begin{equation}
    \Theta = 3H = (2n+1)\,\cfrac{\Dot{X}}{X},
\end{equation}
\begin{equation}
    j=\cfrac{\dddot{a}}{aH^3},
\end{equation}
\begin{equation}
    q=\cfrac{d}{dt}\, \left( \cfrac{1}{H}\right)-1 
\end{equation}
\begin{equation}
    A_h = \frac{1}{3}\, \sum_{i=1}^3 \left(\frac{H_i - H}{H} \right)^2.
\end{equation}
For our model, the average anisotropy will take up constant values, defined by the relation \eqref{AnisotSol}:
\begin{equation}
    A_h =\frac{2(n-1)^2}{(2n+1)^2}.
\end{equation}
We have plotted various interesting parameters as discussed above here.
\begin{figure}[H]
	\begin{center}
\includegraphics[width=0.9\textwidth]{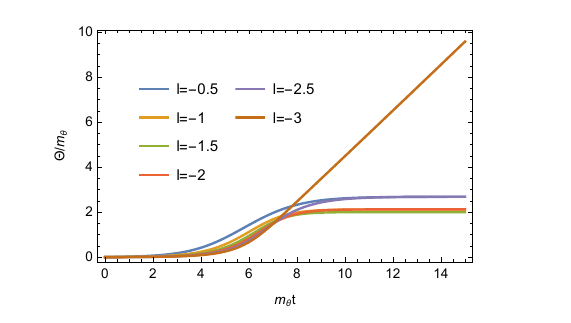}	\end{center}
\caption{The plot of scalar expansion for a given value of l. For $l>-3$ we see scalar expansion approaching constant value, depending on the value of $l$. For $l=-3$ we see it is infinitely growing linearly, which could not correspond to our Universe. }	\label{imp_of_l_scalar_exp}
\end{figure}
\begin{figure}[H]
	\begin{center}
\includegraphics[width=0.9\textwidth]{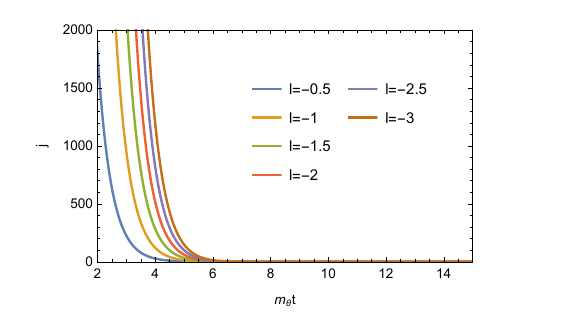}
	\end{center}
\caption{The plot of the jerk parameter for a given $l$. For any value of $l$ it approaches zero.}	\label{imp_of_l_jerk}
\end{figure}
\begin{figure}[H]
	\begin{center}
\includegraphics[width=0.9\textwidth]{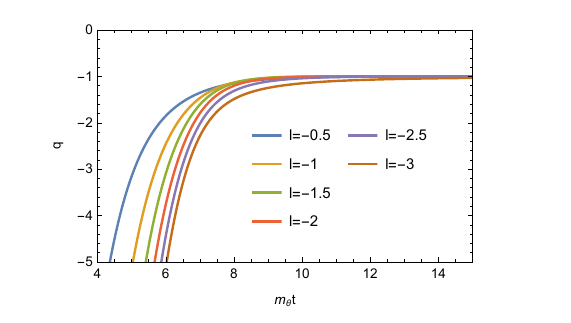}	\end{center}
\caption{The plot of deceleration parameter. We see that for any value of $l$ it is approaching $-1$, which is the case for common $\Lambda$\text{CDM} model. }
	\label{imp_of_l_decel}
\end{figure}
Let us also consider the case $\phi/f \ll 1$, which would allow us to obtain analytical solution for field's equation of motion. This approximation is mathematically the same as in the case in which one would consider scalar field with generic potential.
\\
\\
In this particular case,  equation of motion is as follows
\begin{equation}\label{maineqGen}    \Ddot{\phi}+\cfrac{3\Dot{\phi}^2}{l\phi}+m^2\phi=0,
\end{equation}
which has general solution in the following form
\begin{equation}\label{GenSol}
    \phi(t)=\phi_0\cdot \cos^{\frac{l}{l+3}}{\left(m \sqrt{\frac{l+3}{l}}(l\phi_1 + t) \right)},
\end{equation}
where $\phi_0$ and $\phi_1$ are integration constants.
Solution \eqref{GenSol} leads to average scale factor 
\begin{equation}\label{ScFacGeneric}
    a(t)= \cos^{\frac{1}{l+3}}{\left(m \sqrt{\frac{l+3}{l}}(l\phi_1 + t) \right)}.
\end{equation}
One may notice average scale factor approach infinity in a finite time if $l<-3$, which could indicate a big rip scenario \footnote{However one may still need to do more checks to know if that is indeed the case}. This is also the reason why we do not consider $l<-3$ in our numerical analysis. It is also interesting to note that if $l$ is positive, then this universe has a periodical nature.
With equations \eqref{ScFacGeneric}, \eqref{AnisotSol}, \eqref{AvScFac} we derive metrics' potentials
\begin{multline}\label{metricsPot}
    X(t)=\cos^{\frac{3}{(l+3)(2n+1)}}{\left\{ m (l\phi_1 + t)\sqrt{\frac{l+3}{l}}\right\}},\\
    Y(t)=\cos^{\frac{3n}{(l+3)(2n+1)}}{\left\{ m (l\phi_1 + t)\sqrt{\frac{l+3}{l}}\right\}}.
\end{multline}
Using \eqref{metricsPot}, we derive Kantowski-Sachs cosmological model with homogeneous scalar field and cosmological fluid
\begin{multline}
    ds^2=dt^2-\cos^{\frac{6}{(l+3)(2n+1)}}{\left\{ m (l\phi_1 + t)\sqrt{\frac{l+3}{l}}\right\}}dr^2-\\-\cos^{\frac{6n}{(l+3)(2n+1)}}{\left\{ m (l\phi_1 + t)\sqrt{\frac{l+3}{l}}\right\}}d\Omega^2,
\end{multline}
This could be referred to as the general Kantowski-Sachs model for a generic scalar field.
\section{Conclusion}
\textcolor{black}{In this work we have considered the cosmological dynamics of a scalar field with an axion like potential, in the background of non-standard spacetime metric. The metric we considered was in general that of a inhomogeneous and anisotropic spacetime, which in the far observer limit reduced to the case of a standard Kantowski-Sachs metric. We considered this case in particular, discussing the cosmological dynamics of the scalar field in such a scenario. We analyzed a lot of key parameters in this regard, performing both analytical and numerical analysis to arrive at some interesting findings which we would like to summarize as follows: \begin{itemize}
    \item For the distant observer model, the skewness parameters $ \delta $ and $ \gamma $ approach unity, resulting in zero angular pressures, which leads to a radial alignment of the cosmological fluid's energy density. This anisotropy suggests that the fluid may support cosmic expansion in specific directions while maintaining a dust-like behavior in angular dimensions. 
    \item With the ALP-coupled fluid’s equation of state $ \omega \approx -1 $, the field behaves similarly to dark energy under specific initial conditions. This outcome underscores the potential of the ALP field to simulate effects commonly attributed to dark energy within an anisotropic framework.
    \item The model's predictions include consistent scalar expansion values and an anisotropic distribution of energy density, with a likely impact on cosmic microwave background observations and gravitational wave signatures. These predictions align with general relativity but introduce unique ALP-related anisotropic factors observable for distant observers, thus offering new insights into early universe cosmological models. 
    \end{itemize}}


\section*{Acknowledgements}
\textcolor{black}{We are grateful to Yury Nikolaevich Eroshenko for fruitful discussions. The research by M.K. was carried out in Southern Federal University with financial support of the Ministry of Science and Higher Education of the Russian Federation (State contract GZ0110/23-10-IF).}

\bibliography{references}
\bibliographystyle{unsrt}

\end{document}